\documentclass{ifacconf}

\usepackage{graphicx}      

\usepackage{natbib}        

\usepackage{amssymb}
\usepackage[caption=false,font=normalsize,labelfont=sf,textfont=sf]{subfig}
\usepackage{textcomp}
\usepackage{stfloats}
\usepackage{ragged2e}
\usepackage{xcolor}
\usepackage{amsmath}
\usepackage{float}
\usepackage{amsfonts}
\usepackage[mathscr]{eucal}
\usepackage{longtable}
\usepackage{blindtext}
\usepackage{optidef}
\usepackage{pdflscape}
\usepackage{footnote}
\usepackage{multirow}
\usepackage{soul}
\usepackage{multirow}
\usepackage{tabularx}
\usepackage[markup=underlined]{changes}
\usepackage{flushend}
\usepackage{algorithm,algcompatible}

\algnewcommand\INPUT{\item[\textbf{Input:}]}%
\algnewcommand\OUTPUT{\item[\textbf{Output:}]}%

\newtheorem{remark}{Remark}

\theoremstyle{plain}

\theoremstyle{definition}

\usepackage{nomencl}
\usepackage{float}

\def\BibTeX{{\rm B\kern-.05em{\sc i\kern-.025em b}\kern-.08em
    T\kern-.1667em\lower.7ex\hbox{E}\kern-.125emX}}

\begin{document}
\begin{frontmatter}

\title{\large Path Following Model Predictive Control of a Coupled  Autonomous Underwater Vehicle\thanksref{footnoteinfo}} 

\thanks[footnoteinfo]{This work is supported by the Nigerian Government via the Petroleum Technology Development Fund (PTDF).}

\author[]{Isah A. Jimoh,} 
\author[]{Hong Yue} 

\address[First]{Wind Energy and Control Centre, Department of Electronic and Electrical Engineering, University of Strathclyde,
Glasgow G1 1XW, UK (e-mails: isah.jimoh@strath.ac.uk, hong.yue@strath.ac.uk).}

\begin{abstract}                
The operation of an autonomous underwater vehicle (AUV) faces challenges in following predetermined waypoints due to coupled motions under environmental disturbances. To address this, a 3D path following guidance and control system is developed in this work based on the line-of-sight (LOS) guidance method. Conventionally, the 3D path following problem is transformed into heading and depth control problems, assuming that the motion of the vehicle is decoupled in horizontal and depth coordinates. The proposed control system design avoids this simplifying assumption by transforming the problem into a 3D position and orientation tracking problem. This design is achieved by computing a 2D horizontal coordinate based on the desired heading and then computing a corresponding LOS depth coordinate. A model predictive controller (MPC) is then implemented using the 3D LOS coordinate and the computed orientation vector. The MPC obtains a robust control by solving a minimax optimisation problem considering the effects of unknown ocean disturbances. The effectiveness of the proposed guidance and control system is demonstrated through the simulation of a prototype AUV system. Numerical results show that the AUV can follow predetermined waypoints in the presence of time-varying disturbances, and the system is steered at a constant surge speed that is proportional to the radius of the circle of acceptance used to implement the guidance system. 
 
\end{abstract}


\end{frontmatter}

\section{Introduction}

An autonomous underwater vehicle (AUV) is a marine vehicle capable of independently executing missions through its onboard sensors, guidance and control system. The growing interest in AUVs, both in academia and industry, stems from their potential to reduce risks associated with underwater exploration and resource exploitation \citep{sahoo2019advancements}. The rising importance of consistently monitoring the ocean environment, particularly to address climate change effects, along with the petrochemical industry's shift to seabed platforms, has heightened interest in designing AUVs with enhanced autonomy for tasks like inspection, maintenance, and repair \citep{ribas2015auv}. The increased autonomy required for AUVs to execute tasks without human intervention underscores the necessity for advanced control schemes, which must effectively handle the AUV's nonlinearities, coupling, and environmental disturbances, such as ocean currents and waves, to achieve desired control objectives \citep{jimoh2023autonomous}.

Advances in navigation, guidance and control systems have a significant role in the progress of improving AUV autonomy \citep{zeng2015survey}. Typically, guidance systems and control systems are developed independently. Popular guidance laws such as proportional navigation guidance, Lyapunov-based guidance, and line-of-sight (LOS) guidance were reviewed by \cite{naeem2003review}. In guidance systems, it is a common practice to split the desired AUV path into several paths connected by a set of waypoints that the vehicle needs to go through in order to reach the final destination. Thus, waypoint guidance refers to the process of steering the vehicle from one waypoint to the next \citep{ataei2015three} until the destination is reached. The LOS guidance system is most popular for marine vehicles \citep{fossen2011handbook}.  

Guidance systems based on the LOS strategy are traditionally implemented by generating reference heading between waypoints that are then tracked using a suitable heading controller. This approach applies to marine surface vehicles and AUV horizontal motion control, which has been the main focus of many path-following control strategies \citep{fossen2003line,breivik2004path,fossen2014line}.
For some AUVs whose roll, pitch and heave motions can be assumed decoupled with negligible roll and pitch angles \citep{lamraoui2019path}, their 3D guidance systems can readily be designed using methods similar to those used for marine surface vehicles \citep{ataei2015three}. For the 3D case, in addition to the heading angle, the corresponding depth reference is also determined and tracked to achieve the 3D path following task \citep{khodayari2015modeling}. 
However, these guidance and control schemes cannot be directly applied to achieve accurate waypoint following for AUVs with coupled motion.  Consequently, \cite{lekkas2013line} presented a LOS guidance system for underactuated AUVs with coupled motions, with the assumption that roll motion is negligible. Recently, \cite{zhang2024three} presented findings on coupled motions that take into account roll dynamics, although this was for vehicles following sufficiently smooth paths.

Whereas studies on 3D path following for smooth and continuous trajectories have been reported \citep{zhou2013spatial,yu2017nonlinear,liang2017three,zhang2024three}, there is a lack of investigation on paths described by 3D waypoints. 
In \citep{yao2020path}, a 3D waypoint following MPC design is developed by formulating an error kinematic model, under the assumption of decoupled motions, for predicting the evolution of the linear and angular errors without considering the effects of disturbances.

In this work, a LOS guidance and control system is proposed for coupled AUV with the 3D waypoints following objective. The guidance system makes it possible to achieve the path following at a constant speed, which is preferred for the sake of energy saving in AUV operation \citep{yao2020path}. The MPC control is developed to mitigate AUV velocity fluctuations under environmental disturbance. The proposed scheme is validated via numerical simulation of the fully-actuated Naminow-D AUV \citep{jimoh2024velocity}. The contributions of this work are briefed as follows.
\begin{enumerate}
    \item A LOS guidance system is proposed, which redefines the conventional heading and depth control problems into a 3D LOS path-tracking problem. This approach circumvents the need to develop a kinematic error model. 
    \item An MPC-based control system is developed, aiming to mitigate fluctuations in the AUV velocity vector. This is achieved by employing velocity increment as the optimisation control variable. To ensure robustness against external disturbances, the controller is formulated as a convex minimax control problem, addressing the worst-case scenario of bounded disturbances.
\end{enumerate}
In the rest of this paper,  
Section \ref{sec:problem formulation} describes the kinematic model and the dynamic motion model of the AUV with environmental disturbances. The 3D guidance and control system is presented in Section \ref{sec:3D Guidance System}. Simulation study and results are presented in Section \ref{sec:simulation}. Conclusions and future research directions are discussed in Section \ref{sec:conclusions}.

\section{Coupled AUV Model}\label{sec:problem formulation}

A two reference frame system, the body-fixed and the inertia reference frames, is generally used to describe the motion of an AUV. For coordinate transformation between these two frames, the following kinematic model is used:
\begin{equation}\label{eqn:kinematics}
   {\boldsymbol{\dot{\eta}} =\mathbf{J}\boldsymbol{({\eta})\nu}},
    \end{equation}
where ${\boldsymbol{\eta}} = \left[x\ y\ z\ \phi\ \theta\ \psi \right]^\top$ is the vector of linear and angular positions and $\boldsymbol{\nu}=\left[u\ v\ w\ p\ q\ r \right]^\top$ is the vector of the linear and angular velocities of the AUV. Here, $x,\ y,\ z$  denote the spatial coordinates in the 3D Cartesian coordinate; $\phi,\ \theta$ and $\psi$ are the roll angle, pitch angle and yaw angle, respectively; $u$, $ v$ and $w$ are the linear velocities and $p,\ q$ and $r$ are the angular velocities in the body-fixed frame; $\mathbf{J}\boldsymbol{({\eta})}\in\mathbb{R}^{6\times 6}$ is the rotation matrix given by  
\begin{equation}\nonumber
        \mathbf{J}\boldsymbol{({\eta})}= \begin{bmatrix} \mathbf{J_1\boldsymbol{({\eta})}}&  \mathbf{0_{3\times 3}}\\ \mathbf{0_{3\times 3}}& \mathbf{J_2\boldsymbol{({\eta})}} \end{bmatrix},
    \end{equation}

     \begin{equation}\nonumber
     \begin{aligned}         
        \mathbf{J_1\boldsymbol{({\eta})}} =&\left[\begin{matrix}\mathrm{cos}\theta \mathrm{cos}\psi &-\mathrm{sin}\psi \mathrm{cos}\phi  +  \mathrm{cos}\psi \mathrm{sin}\theta\mathrm{sin}\phi \\
        \mathrm{sin}\theta \mathrm{cos}\psi &\mathrm{cos}\psi \mathrm{cos}\phi  +  \mathrm{sin}\phi \mathrm{sin}\theta\mathrm{sin}\psi\\
        -\mathrm{sin}\theta &\mathrm{cos}\theta \mathrm{sin}\phi \end{matrix} \right. \\ &
       \left. \begin{matrix} \mathrm{cos}\psi \mathrm{cos}\phi\mathrm{sin}\theta + \mathrm{sin}\psi \mathrm{sin}\phi\\
         \mathrm{sin}\theta \mathrm{sin}\psi\mathrm{cos}\phi - \mathrm{cos}\psi \mathrm{sin}\phi\\
         \mathrm{cos}\theta \mathrm{cos}\phi
        \end{matrix} \right],
     \end{aligned}    
    \end{equation}
    \begin{equation}\nonumber
        \mathbf{J_2}\boldsymbol{({\eta})}= \begin{bmatrix}
        1& \mathrm{sin}\phi\mathrm{tan}\theta & \mathrm{cos}\phi\mathrm{tan}\theta\\
        0&\mathrm{cos}\phi & -\mathrm{sin}\phi\\
        0&\mathrm{sin}\phi/\mathrm{cos}\theta & \mathrm{cos}\phi/\mathrm{cos}\theta
        \end{bmatrix}.
    \end{equation}
    

The 6-degree-of-freedom (DoF) AUV model with environmental disturbances is given by
    \begin{equation}\label{eqn:dynamics}
     \mathbf{M}\dot{\boldsymbol{\nu}} +  \mathbf{C}(\boldsymbol{\nu})\boldsymbol{\nu} + \mathbf{D}(\boldsymbol{\nu})\boldsymbol{\nu} + \mathbf{g}(\boldsymbol{\eta}) = \boldsymbol{\tau}+\boldsymbol{\tau}^w, 
    \end{equation} 
where $\mathbf{M}\succ \mathbf{0}\in\mathbb{R}^{6\times 6}$ is the AUV's inertia matrix that comprises of the rigid body and added mass components, $\mathbf{C}\boldsymbol{({\nu})}\in\mathbb{R}^{6\times 6}$ is the Coriolis-centripetal matrix, $\mathbf{D}\boldsymbol{({\nu})}\in\mathbb{R}^{6\times 6}$ is the hydrodynamic damping matrix and $\mathbf{g}\boldsymbol{({\eta})}\in\mathbb{R}^{6}$ is the AUV's weight and buoyancy forces vector. Furthermore, $\boldsymbol{\tau}= [\tau_X\ \tau_Y\ \tau_Z\ \tau_K\ \tau_M\ \tau_N]^\top\in\mathcal{T}\subset\mathbb{R}^{6}$ denotes the generated constrained input forces and moments driving the AUV from its centre of gravity, with the constraint set $\mathcal{T}$ defined as
\begin{equation}
    \mathcal{T}:=\left\{\boldsymbol{\tau}\subset \mathbb{R}^{6}: |\boldsymbol{\tau}|\le \boldsymbol{\bar \tau}  \right\}
\end{equation}
where $\boldsymbol{\bar \tau}$ denotes the bounds on the control inputs.
The unknown ocean waves disturbance is represented by $\boldsymbol{\tau}^w=[\tau^w_X\ \tau^w_Y\  \tau^w_Z\ 0\ 0\  0]^\top\in\mathbb{R}^{6}$ that affects the AUV's motion.  
 
 The ocean waves is modelled as \citep{fossen2011handbook}:
 \begin{equation}\label{eqn:wavemodel-1}
   \begin{bmatrix}\dot{z}^w_{i,1}\\ \dot{z}^w_{i,2}       
   \end{bmatrix} = \begin{bmatrix}
       0 & 1 \\
       -\omega^2_{0,i} & -2\xi_{i}\omega_{0,i}
   \end{bmatrix}\begin{bmatrix}{z}^w_{i,1}\\ {z}^w_{i,2}       
   \end{bmatrix} + \begin{bmatrix}0 \\ K_{w,i}       
   \end{bmatrix}w_i
\end{equation}
 \begin{equation}\label{eqn:wavemodel-2}
         {\tau}^w_{i} = \begin{bmatrix}
             0&1
         \end{bmatrix}\begin{bmatrix}{z}^w_{i,1}\\ {z}^w_{i,2}       
   \end{bmatrix} + d_i
     \end{equation} 
in which the subscript $i$ ($=X,\ Y,\ Z$) {corresponds to the three line directions}, the amplitude of ${\boldsymbol{\tau}}^w_{i}$ in the $i-$th direction can be adjusted by the parameter $K_{w,i}$. The term $w_i$ is a zero-mean white noise, $\xi_i$ is the damping coefficient, $\omega_{0,i}$ is the wave peak frequency.
  The variables  $d_i$ are modelled as slowly varying bias terms bounded by $|d_i|\le d_{\mathrm{max}}$.


\section{Guidance and Control System}
\label{sec:3D Guidance System}
\subsection{Line-of-Sight (LOS) Guidance System}
  Due to its simplicity and ease of implementation, the LOS method is used by most guidance laws \citep{naeem2003review}. 
  In 3D motion control, the angles $\phi$ and $\theta$ and the corresponding angular velocities $p$ and $q$ are often assumed to be negligible. This assumption applies mostly for underactuated AUVs. This work avoids such assumptions by directly determining the LOS coordinates and the orientation vector needed for the desired waypoints following task.
  
\begin{figure}
    \centering
    \includegraphics[width=\linewidth]{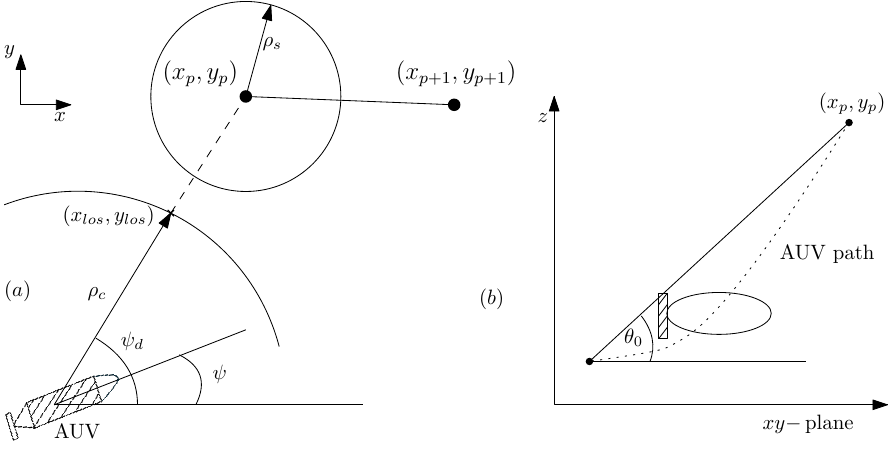}
    \caption{LOS guidance system for waypoints following}
    \label{fig:LOS_guidance}
\end{figure}  
With the 2D horizontal plane, as illustrated in Fig. \ref{fig:LOS_guidance}(a), given the current AUV position as $(x(k),y(k))$ at time $k$, the desired heading angle is given as \citep{naeem2003review}:
\begin{equation}\label{eqn:desired_angle}
    \psi_d=\text{tan}^{-1}\left(\frac{y_p-y(k)}{x_p-x(k)} \right),
\end{equation}
where $(x_p,y_p)$ with $p=1,2,...,m$ are the coordinates of the $m$ waypoints in the horizontal plane. Since the waypoint position may be significantly distant from the AUV position, the LOS coordinate $(x_{los}(k),y_{los}(k))$ is defined by the so-called circle of acceptance as follows:
\begin{subequations}\label{eqn:los_guidance}
\begin{equation}\label{eqn:los_replanninga}
  \begin{aligned}
    \left(x_{los}(k)-x(k)\right)^2  + \left(y_{los}(k)-y(k)\right)^2 = \rho_c^2,\\
    \end{aligned}
    \end{equation}
    \begin{equation}\label{eqn:los_replanningb}
        \left(y_{los}(k) - y(k)\right) =\left(x_{los}(k) - x(k) \right) \text{tan}\psi_d.\\
        \end{equation}
\end{subequations}
Here $\rho_c$ is the radius of the circle of acceptance. It is evident that the desired yaw angle $\psi_d$ will be maintained by the LOS coordinates obtained from (\ref{eqn:los_guidance}) and the LOS point will lie on the circle of acceptance.  

The value of $\rho_c$ directly impacts the speed of the AUV along the prescribed horizontal path since it determines the average distance between successive LOS coordinates. To see this, from  (\ref{eqn:los_replanninga}) and (\ref{eqn:los_replanningb}), there is
\begin{equation}\label{eqn:los_speed1}
        x_{los}(k) =x(k)  \pm\frac{\rho_c}{(1+\mathrm{tan}^2\psi_d)^{1/2}}.
        \end{equation}
In (\ref{eqn:los_speed1}), the positive sign corresponds to $(x_p-x(k)) \ge 0$ while $(x_p-x(k))<0$ is related to the negative sign. Subtract $x_{los}(k-1)$ from both sides of (\ref{eqn:los_speed1}) and then divide by $T_s$ to obtain:
        \begin{equation}\label{eqn:los_speed2}
        u_{los}(k)=\frac{x(k)-x_{los}(k-1)}{T_s}  \pm\frac{\rho_c}{T_s(1+\mathrm{tan}^2\psi_d)^{1/2}},
        \end{equation}
where $u_{los}(k)=\frac{x_{los}(k)-x_{los}(k-1)}{T_s}$ represents the discrete-time approximation of the LOS surge speed. 
This equation reveals that the LOS surge speed is directly influenced by the circle of acceptance used.
When following a straight-line path where \( \psi_d \) remains constant, the speed is mainly dictated by \( \rho_c \). 
A significant speed change may occur when the vehicle adjusts its heading $\psi_d$ to track the next waypoint $(x_{p+1},y_{p+1})$. Indeed, opting for \( \rho_c < L \) ensures that the AUV maintains a relatively low speed along the path, which is often desirable for underwater tasks, as it limits the effects of coupling.

Since the target LOS depth is not necessarily equal to the desired depth, 
we compute the LOS depth to be proportional to the distance between the LOS horizontal coordinate and the AUV position. 
Consider the angle $\theta_0$, defined by the line joining $z_p$ on the $z-$axis to the waypoint $(x_p,y_p)$ on the horizontal plane, as shown in Fig. \ref{fig:LOS_guidance}(b). This angle is computed as
\begin{equation}
    \theta_0=\text{tan}^{-1}\frac{z_p-z(k)}{\sqrt{(x_p-x(k))^2+(y_p-y(k))^2}}.
\end{equation}
To keep  this angle for any current depth $z(k)$, the desired depth of AUV is given by
\begin{equation}\label{eqn:gs_z1}
z_{los}=z(k)+\text{tan}\theta_0 \cdot \rho_c
\end{equation}
Aside from the 3D LOS path to be followed, it is important to define the condition necessary for switching from one waypoint, $(x_p,y_p,z_p)$, to follow to the next waypoint, $(x_{p+1},y_{p+1},z_{p+1})$. This is achieved by defining the ``sphere of acceptance $\rho_s$" around each waypoint in the 3D environment (see Fig. \ref{fig:LOS_guidance}). The $(x_{p+1},y_{p+1},z_{p+1})$ is chosen when the inequality is satisfied.
\begin{equation}\label{eqn:los_switching_cond}
     \begin{aligned}
     \left(x_p-x(k)\right)^2 & +  \left(y_p-y(k)\right)^2 +  \left(z_p-z(k)\right)^2 \le \rho^2_s.
\end{aligned}
\end{equation}
The sphere of acceptance can be taken to be twice of the length of the vehicle \citep{fossen2011handbook}. 

Furthermore, there is a need to define the reference orientation for the AUV while moving along the path. Consider the straight line path 
linking the two waypoints $(x_p,y_p,z_p)$ and  $(x_{p+1},y_{p+1},z_{p+1})$. The yaw and pitch angles from the path coordinates at $(x_p,y_p,z_p)$ with respect to the inertia frame is given as
\begin{equation}
    \begin{aligned}
    \psi_p &= \mathrm{arctan2}\left(y_{p+1}-y_{p},x_{p+1}-x_{p}\right) \\
    \theta_p &= -\mathrm{tan}^{-1}\frac{z_{p+1}-z_{p}}{\sqrt{(x_{p+1}-x_{p})^2+(y_{p+1}-y_{p})^2}}
\end{aligned}
\end{equation}
where $\mathrm{arctan2}(\cdot)$ is the four quadrant  inverse function used to ensure $-\pi\le \psi_p\le\pi$. 
The {yaw and pitch} angles are used to define the desired AUV orientation along the straight-line path when the roll angle is set to zero as roll motion needs to be kept minimum. Hence, the reference signal to be tracked to achieve path following is given by
\begin{equation}
    \mathbf{y}^{\text{ref}}(k) = [x_{los}(k)\ y_{los}(k)\ z_{los}(k)\ 0\ \theta_p\ \psi_p]^\top.
\end{equation}
The reference $\mathbf{y}^{\text{ref}}(k) $ is used in the controller design next.

\subsection{Path Tracking Control Based on Mini-Max MPC}

The kinematic model (\ref{eqn:kinematics}) is discretised into the following model with sampling time $T_s$:
\begin{equation}\label{eqn:kin_dis}
    \boldsymbol{\eta}(k+1) = \boldsymbol{\eta}(k) + \mathbf{J}(k)\boldsymbol{\nu}(k),
\end{equation}
where $\mathbf{J}(k)= \mathbf{J}(\boldsymbol{{\eta}}(k))T_s.$    
   The AUV increment velocity is defined as
    \begin{equation}\label{eqn:velocity_increment}
        \Delta\boldsymbol{\nu}(k)= \boldsymbol{\nu}(k)-\boldsymbol{\nu}(k-1).
    \end{equation}    
The state-space model is obtained by combining (\ref{eqn:kin_dis}) and (\ref{eqn:velocity_increment}) as follows:
\begin{equation}\label{eqn:ss_system}
    \begin{aligned}
    \mathbf{x}(k+1) &= \mathbf{A}(\boldsymbol{\eta}(k))\mathbf{x}(k) + \mathbf{B}(\boldsymbol{\eta}(k))\Delta\boldsymbol{\nu}(k),\\
        \mathbf{y}(k) &= \mathbf{G}\mathbf{x}(k),        
    \end{aligned}
\end{equation}
in which $\mathbf{x}(k)= [{\boldsymbol{\eta}}(k)^\top\ {\boldsymbol{\nu}}(k)^\top]^\top$, $\mathbf{y}(k)={\boldsymbol{\eta}}(k)$, $\mathbf{G}=[\mathbf{I}_6\ \mathbf{0}_6]$, and
\begin{equation}\label{eqn:A_and_B}
  \mathbf{A}(\boldsymbol{\eta}(k))=\begin{bmatrix}
   \mathbf{I} &\mathbf{J}(k) \\
   \mathbf{0} &\mathbf{I}
    \end{bmatrix},\ \mathbf{B}(\boldsymbol{\eta}(k))=\begin{bmatrix}
   \mathbf{J}(k) \\\mathbf{I}
    \end{bmatrix}.
\end{equation}
Here $ \mathbf{I}$ and $ \mathbf{0}$ represent identity and zero matrices of appropriate dimensions, respectively.

Define the stacked versions of the predicted state, output and velocity increment as
\begin{equation}\label{eqn:inputdef}
\begin{aligned}
\mathbf{X}(k) &= \begin{bmatrix}
 \mathbf{x}({k+1|k})^\top& \ldots &\mathbf{x}({k+N|k})^\top
\end{bmatrix}^\top,\\
\mathbf{Y}(k) &=\begin{bmatrix}
 \mathbf{y}({k+1|k})^\top& \ldots &\mathbf{y}({k+N|k})^\top
\end{bmatrix}^\top,\\
    \mathbf{U}(k) &= \begin{bmatrix} \Delta\boldsymbol{ \nu}({k|k})^\top & \ldots &\Delta\boldsymbol{ \nu}({k+N-1|k})^\top\end{bmatrix}^\top,\
\end{aligned}
\end{equation}
where $N$ is the prediction horizon, then the state prediction model can be written in a compact form as
\begin{equation}\label{eqn:pred_model}
    \mathbf{X}(k)= \Tilde{\mathbf{A}}\mathbf{x}(k|k)+\Tilde{\mathbf{B}}\mathbf{U}(k),
\end{equation}
\begin{figure*}[ht]
    \centering
    \includegraphics[width=0.7\linewidth]{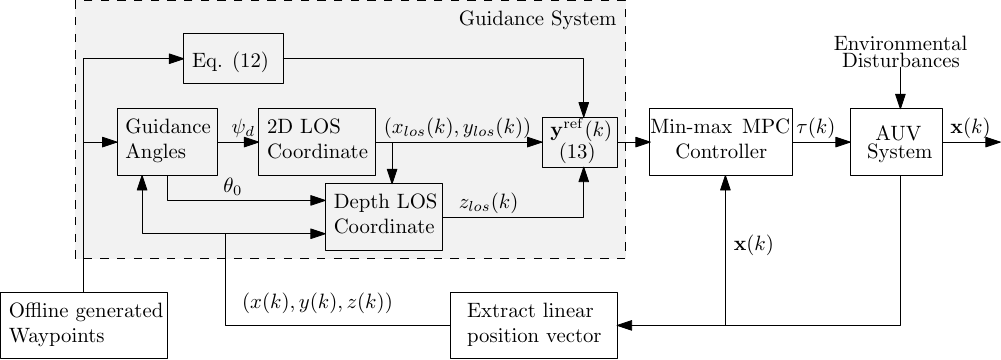}
    \caption{Proposed 3D path following control system for coupled AUVs}    \label{fig:Proposed_LOS_GS}
\end{figure*}
where  $\Tilde{\mathbf{A}}$ and $\Tilde{\mathbf{B}}$ are structured using \eqref{eqn:A_and_B} to obtain the predictions over the horizon \citep{zhang2019mpc}.
Given that $\boldsymbol{\eta}({k+j|k})$ denotes the $j$-th prediction of $\boldsymbol{\eta}$ at time $k$, the system matrices are assumed to be constant over the prediction horizon $N$, that is, $\mathbf{A}(\boldsymbol{\eta}({k+j|k})=\mathbf{A}(\boldsymbol{\eta}(k))$ and $\mathbf{B}(\boldsymbol{\eta}({k+j|k})=\mathbf{B}(\boldsymbol{\eta}(k))$ for all $j= 1,\ldots, N$. This means that the model is inherently treated as a linear-time varying system that is updated at each sampling time step. This reduces the computational load as the need for an update at every prediction step $j$ is avoided.  

Denote $\mathbf{d}(k)\in\mathcal{D}\subset \mathbb{R}^{12}$ as the lumped term to include environmental disturbances and the uncertainties in the state predictions, 
defined by
\begin{equation}
    \mathcal{D}:=\left\{\mathbf{d}(k)\subset \mathbb{R}^{12}: |\mathbf{d}(k)|\le \Bar{\mathbf{d}}  \right\}
\end{equation}
where $\Bar{\mathbf{d}}$ represents the upper bounds on $\mathbf{d}(k)$. 
The state space representation of the AUV model becomes
\begin{equation}\label{eqn:ss_system1}
    \mathbf{x}(k+1) = \mathbf{A}(\boldsymbol{\eta}(k))\mathbf{x}(k) + \mathbf{B}(\boldsymbol{\eta}(k))\Delta\boldsymbol{\nu}(k) + \mathbf{d}(k).
\end{equation}

Let $\mathbf{D}(k)=[\mathbf{d}({k+1|k})^\top\ \ldots\ \mathbf{d}({k+N|k})^\top]^\top$, the prediction model (\ref{eqn:pred_model}) with additive external disturbance yields
\begin{equation}\label{eqn:pred_model_dis}
    \mathbf{X}(k)= \Tilde{\mathbf{A}}\mathbf{x}(k|k)+\Tilde{\mathbf{B}}\mathbf{U}(k)+\mathbf{D}(k).
\end{equation}
The output prediction of the AUV system is given by
\begin{equation}\label{eqn:output_pred}
    \mathbf{Y}(k) = \mathbf{\tilde G}\mathbf{X}(k),
\end{equation}
with $\mathbf{\tilde G} = \mathrm{diag}(\mathbf{G},\ldots,\mathbf{G})$. With the upper bounds known on the disturbances, the objective of robust control is to minimise the worst case scenario, \textit{i.e.,} minimise the tracking error under the maximum uncertainties subject to input constraints. 

The objective function of the finite horizon constrained optimal control problem (FHCOCP) is defined to minimise the path {following} error which is the difference between the AUV position and {the reference} $\mathbf{y}^{\text{ref}}$ as 
\begin{equation}\label{eqn:cost_fun_conf}
\begin{aligned}
    V\left(\mathbf{U}(k),\mathbf{Y}(k)\right) =    &\sum^{N}_{j=1}\parallel\mathbf{y}({k+j|k})-\mathbf{y}^{\text{ref}}({k+j|k})\parallel^2_{\mathbf{Q}}\\    
    +&\sum^{N_u-1}_{j=0}\parallel \Delta\boldsymbol{\nu}({k+j|k}) \parallel^2_{\mathbf{R}},
\end{aligned}
\end{equation}
where $\mathbf{Q}\in\mathbb{R}^{6\times6}$ and $\mathbf{R}\in\mathbb{R}^{6\times6}$ are positive definite matrices, and $N_u (N_u<N)$ is the control horizon in MPC. The control input sequence is  $\mathbf{U}(k)=\{\Delta{\boldsymbol{\nu}}({k|k}),  \ldots, \Delta{\boldsymbol{\nu}}({k+N_u-1|k})\}$. The control law can be structured by assuming that there is no variation in the control signal beyond $N_u$ \citep{fernandez2007model}, \textit{i.e.,} $\Delta{\boldsymbol{\nu}}({k+j|k})=\mathbf{0}$ for $j=N_u,\ldots,N-1$. In this case, the dimension of the MPC problem is reduced from $6N$ to $6N_u$. 
It is pertinent to note that by taking the velocity increment as the control variables in optimisation, the variations in the vehicle's speed are minimised compared to using the velocity term directly.
Employing (\ref{eqn:cost_fun_conf}) as the performance index, the FHCOCP  is formulated as
\begin{equation}\label{eqn:minimax_cost} 
  \begin{aligned}
 \mathbf{U}(k)^* = &\ \arg \underset{\mathbf{U(k)}}{\mathrm{min}}\  \underset{{\mathbf{D}}(k)}{\mathrm{max}} \ V\left(\mathbf{U}(k),\mathbf{Y}(k)\right)\\
  \mathrm{s. \ t.:} \
  &  (\ref{eqn:pred_model_dis})\ \& \ (\ref{eqn:output_pred}),\\
  & {\boldsymbol\tau}({k+j|k})\in \ \mathcal{T},\ j\ge 0,\\ 
 &   {\mathbf{x}}({k|k})=\ {\mathbf{x}}(k).\\
\end{aligned}
\end{equation}
 To enforce constraints on input forces and moments, the variable ${\boldsymbol\tau}({k+j|k})$ are expressed in terms of the control variables $\Delta{\boldsymbol{\nu}}(k+j|k)$. Hence, we re-write (\ref{eqn:dynamics}) in the compact form
      \begin{equation}\label{eqn:inputgeneralised}
     \boldsymbol{\tau}=\mathbf{{M}\dot{\boldsymbol{\nu}} + \chi(\boldsymbol{\nu})}, 
    \end{equation}
    in which  $\chi(\boldsymbol{\nu}) =  \mathbf{C}(\boldsymbol{\nu})\boldsymbol{\nu} + \mathbf{D}(\boldsymbol{\nu})\boldsymbol{\nu}+\mathbf{g}(\boldsymbol{\eta})$. The unknown disturbance  $\boldsymbol{\tau}^w$ is not included here.
    
    Given the past measured velocities $\boldsymbol{\nu}(k-1)$, the discrete-time approximation of (\ref{eqn:inputgeneralised}) can be employed to calculate the inputs at $k$. This is based on the computed optimal vehicle velocity increment, $\Delta\boldsymbol{\nu}(k)^*=\Delta\boldsymbol{\nu}({k|k})^*$, and is expressed as follows:
    \begin{equation}\label{eqn:final_control}
     \begin{aligned}
     \boldsymbol{\tau}(k) =\bar{\mathbf{M}}\Delta\boldsymbol{\nu}(k)^*  + \chi(\boldsymbol{\nu}(k-1)),
     \end{aligned}
    \end{equation}
    where $T_s\bar{\mathbf{M}} = \mathbf{M}$.
    The constraints on the generalised input forces {and moments} can be enforced over the control horizon, that is, $j = 0,\ldots, N_u-1$, using (\ref{eqn:final_control}), as 
    \begin{equation}\label{eqn:final_control_pred}
     \begin{aligned}
     \boldsymbol{\tau}({k+j|k}) =\bar{\mathbf{M}}\Delta\boldsymbol{\nu}({k+j|k})  + \chi(\boldsymbol{\nu}(k-1)).
     \end{aligned}
    \end{equation}

\begin{remark}
The convex optimisation problem (\ref{eqn:minimax_cost}) can be readily solved using standard {optimisation} techniques. Compared to the formulation by \cite{yan2012model}, which leads to nonlinear optimisation, our formulation (\ref{eqn:minimax_cost}) is quadratic because the system $(\mathbf{A}(\boldsymbol{\eta}(k)),\mathbf{B}(\boldsymbol{\eta}(k)))$ is linear parameter-dependent.  
\end{remark}

The proposed configuration of the guidance and control system is illustrated in Fig. \ref{fig:Proposed_LOS_GS}.

\section{Simulation Study}\label{sec:simulation}
The simulation study considers the dynamic model of the Naminow-D AUV whose parameters are described in \citep{jimoh2024velocity}. The length of the AUV including installed sensors is $3.0$m. The control task is for the guidance system to guide the AUV to follow a set of waypoints which have been determined offline. The input forces are constrained to be $\pm 2,000$ N, \textit{i.e.}, $\boldsymbol{\bar\tau}=2,000$. The simulation study is conducted in MATLAB environment. The disturbances are assumed to be bounded by $\mathbf{\bar d}\le 0.5$. The tuned parameters for the guidance and control system are given as follows:  $\mathbf{Q}=\mathrm{diag}(5, 5, 5, 0.1, 0.1, 0.1)$, $\mathbf{R}=18\mathbf{I}$, $N_u=2$, $N=10$, $\rho_c=L/6=0.5$m and $\rho_s=L$.

\begin{figure}[H]
    \centering
    \includegraphics[width=\linewidth, trim = 4.8cm 10.90cm 5.0cm 11cm, clip]{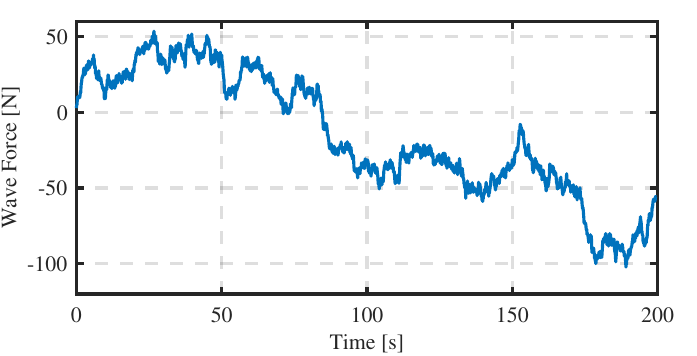}
    \caption{Wave disturbances affecting the AUV's motion.}
    \label{fig:wave_force}
\end{figure}

The ocean waves are modelled using (\ref{eqn:wavemodel-1})-(\ref{eqn:wavemodel-2}) with $\xi_i=0.2573, \omega_{e,i}=0.8\ \text{rad/s}, K_{w,i}=1.5$, $w_i$ is a zero-mean white noise with standard deviation of $0.15$ and $d_i$ is modelled as a Wiener process that lie in the range $[-100,100]$. 
The ocean waves disturbance is modelled such that $\boldsymbol{\tau}^w_X=\boldsymbol{\tau}^w_Y=\boldsymbol{\tau}^w_Z$ and the time profile is shown in Fig. \ref{fig:wave_force}. 

The three waypoints are taken as the scaled version of the optimally generated path in \citep{ataei2015three},   $A=[20,\ 40,\ -16]$, $B=[50,\ 20,\ -16]$ and $C=[70,\ 50,\ -8]$. The destination is located at $[40,\ 70,\ -4].$
These waypoints represent a complicated path, making it suitable to test the effectiveness of a guidance system \citep{ataei2015three}.

\begin{figure*}[h] 
    \centering
  \subfloat[\label{fig:3D_dist}]{%
       \includegraphics[width=.5\linewidth, trim = 3.5cm 8.7cm 3.5cm 8.8cm, clip]{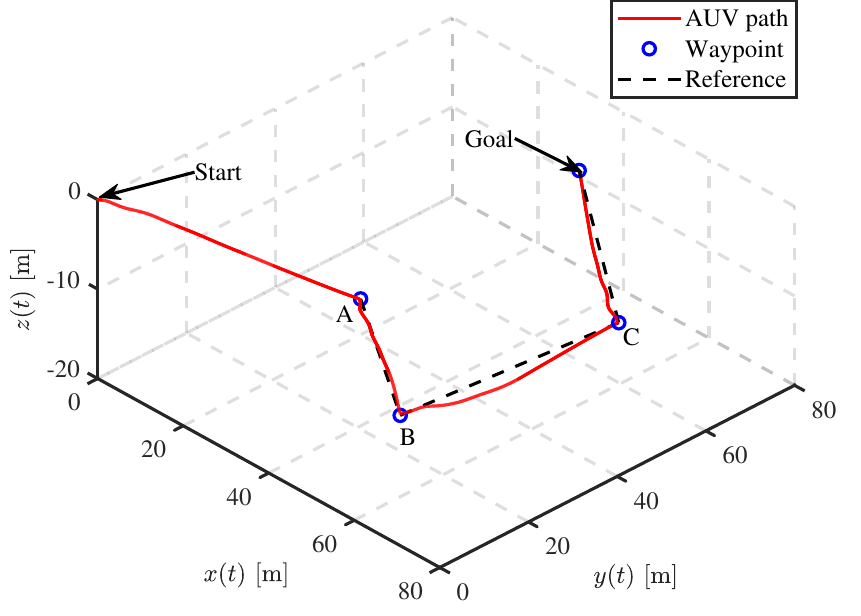}}
    \hfill
  \subfloat[\label{fig:2D_dist}]{%
        \includegraphics[width=.5\linewidth, trim = 5cm 9.5cm 5cm 10cm, clip]{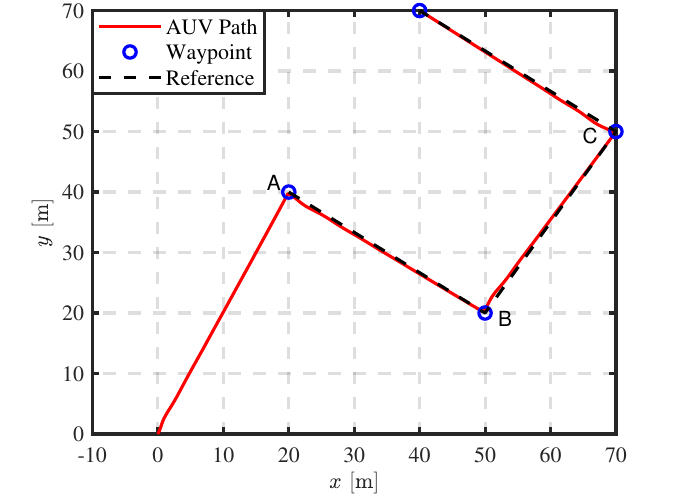}}
    \\
  \subfloat[\label{fig:velocity_dist}]{%
        \includegraphics[width=.5\linewidth, trim = 5cm 9.5cm 5cm 9.5cm, clip]{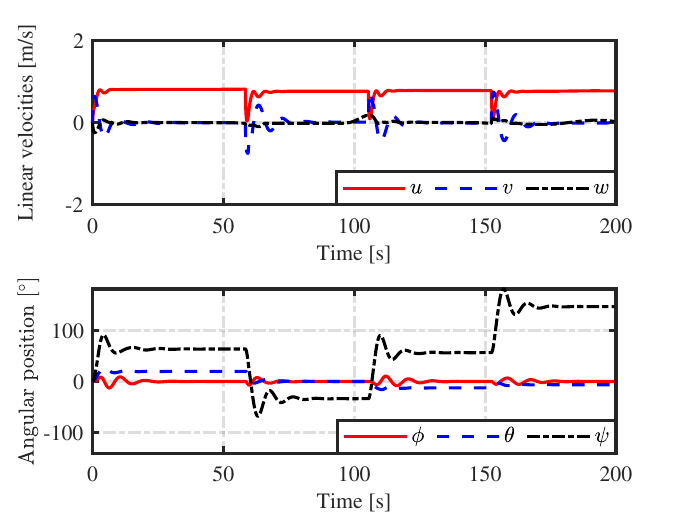}}
    \hfill
  \subfloat[\label{fig:inputs_dist}]{%
        \includegraphics[width=.5\linewidth, trim = 5cm 9.5cm 5cm 9.5cm, clip]{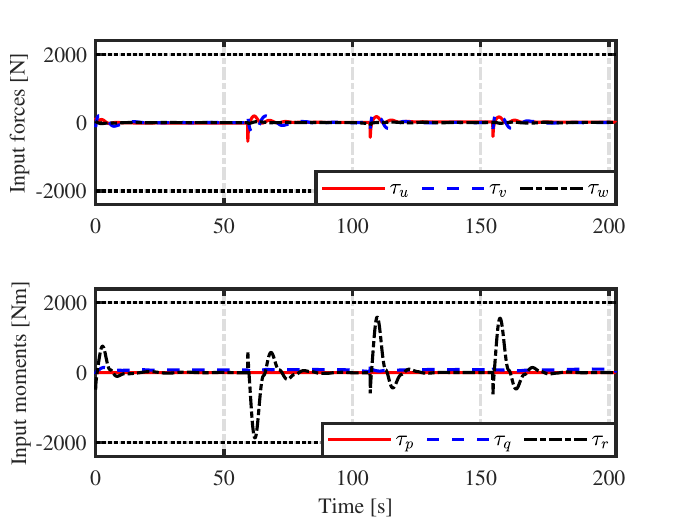}}
  \caption{(a) 3D path following task; (b) Path following task in the horizontal plane; (c) Linear and angular velocities of the AUV; (d) Input forces and moments of the AUV }
  \label{fig:PFC_conference_mm} 
\end{figure*}

The results for the investigated scenario are presented in Fig. \ref{fig:PFC_conference_mm}. It is seen that the AUV can satisfactorily go through the prescribed waypoints as shown in Fig. \ref{fig:3D_dist} {and Fig. 4b} despite the effects of environmental disturbances.
 The velocities are shown in Fig. \ref{fig:velocity_dist}.  A point worthy of note is that the surge velocity is maintained at an approximately constant level of $u\approx 0.8$ m/s. Although not shown, a surge speed of $1$ m/s is obtained when $\rho_c=L/4$ is used in the simulation. In Fig. \ref{fig:velocity_dist}, it is also seen that the angular positions, especially the yaw angle, are adaptively functioned to adjust the AUV orientation as it navigates through the waypoints. The roll motion is stabilised by minimising the roll angle of the AUV such that the average roll angle is $0.015^\circ$ with a maximum absolute value of $12.5^\circ$. Keeping the roll angle  close to zero is desirable for the AUV's stability \citep{fossen2011handbook}. 
The waypoint following is achieved with the input constraints satisfied as seen in Fig. \ref{fig:inputs_dist}. Interestingly, input forces and moments are minimal except during orientation changes.

 \section{Conclusions and Future Works}\label{sec:conclusions}
A guidance and control system for a coupled AUV has been proposed in this study. The system is designed to enable an AUV to follow 3D waypoints in the presence of unknown but bounded environmental disturbances. The proposed guidance system operates without relying on assumptions of negligible roll motion or decoupled motion. It uses the desired heading angle to determine the LOS horizontal coordinate, and the depth reference is proportional to the distance between the AUV's position and the LOS position in the horizontal plane. The AUV's orientation along the straight-line path between two consecutive waypoints is defined, and a MPC control system {is developed to} regulate position and orientation errors for path following. The control system addresses disturbances by formulating a minimax optimisation problem, considering worst-case scenarios based on the disturbance upper bounds. Simulation results demonstrate {the robustness of} the proposed guidance and control system in guiding the AUV through waypoints despite environmental disturbances. 
Future research would consider real-time collision avoidance since the ocean environment can only be partially known during waypoints generation. Further studies are required to explore control system stability and experimental validation. 

\bibliography{references}

\begin{thebibliography}{22}
\providecommand{\natexlab}[1]{#1}
\providecommand{\url}[1]{\texttt{#1}}
\providecommand{\urlprefix}{URL }
\expandafter\ifx\csname urlstyle\endcsname\relax
  \providecommand{\doi}[1]{doi:\discretionary{}{}{}#1}\else
  \providecommand{\doi}{doi:\discretionary{}{}{}\begingroup \urlstyle{rm}\Url}\fi

\bibitem[{Ataei and Yousefi-Koma(2015)}]{ataei2015three}
Ataei, M. and Yousefi-Koma, A. (2015).
\newblock Three-dimensional optimal path planning for waypoint guidance of an autonomous underwater vehicle.
\newblock \emph{Rob. Auton. Syst.}, 67, 23--32.

\bibitem[{Breivik and Fossen(2004)}]{breivik2004path}
Breivik, M. and Fossen, T.I. (2004).
\newblock Path following of straight lines and circles for marine surface vessels.
\newblock \emph{IFAC-PapersOnLine}, 37(10), 65--70.

\bibitem[{Fernandez-Camacho and Bordons(2007)}]{fernandez2007model}
Fernandez-Camacho, E. and Bordons, C. (2007).
\newblock \emph{Model Predictive Control}.
\newblock Springer, second edition.

\bibitem[{Fossen(2011)}]{fossen2011handbook}
Fossen, T.I. (2011).
\newblock \emph{Handbook of Marine Craft Hydrodynamics and Motion Control}.
\newblock John Wiley \& Sons.

\bibitem[{Fossen et~al.(2003)Fossen, Breivik, and Skjetne}]{fossen2003line}
Fossen, T.I., Breivik, M., and Skjetne, R. (2003).
\newblock Line-of-sight path following of underactuated marine craft.
\newblock \emph{IFAC-PapersOnLine}, 36(21), 211--216.

\bibitem[{Fossen et~al.(2014)Fossen, Pettersen, and Galeazzi}]{fossen2014line}
Fossen, T.I., Pettersen, K.Y., and Galeazzi, R. (2014).
\newblock Line-of-sight path following for dubins paths with adaptive sideslip compensation of drift forces.
\newblock \emph{IEEE Trans. Control Syst. Technol.}, 23(2), 820--827.

\bibitem[{Jimoh et~al.(2023)Jimoh, Yue, and K{\"u}{\c{c}}{\"u}kdemiral}]{jimoh2023autonomous}
Jimoh, I.A., Yue, H., and K{\"u}{\c{c}}{\"u}kdemiral, I.B. (2023).
\newblock Autonomous underwater vehicle positioning control-a velocity form {LPV-MPC} approach.
\newblock \emph{IFAC-PapersOnLine}, 56(2), 4388--4393.

\bibitem[{Jimoh and Yue(2024)}]{jimoh2024velocity}
Jimoh, I.A. and Yue, H. (2024).
\newblock A velocity form model predictive control of an autonomous underwater vehicle.
\newblock \emph{Authorea Preprints}.

\bibitem[{Khodayari and Balochian(2015)}]{khodayari2015modeling}
Khodayari, M.H. and Balochian, S. (2015).
\newblock Modeling and control of autonomous underwater vehicle ({AUV}) in heading and depth attitude via self-adaptive fuzzy {PID} controller.
\newblock \emph{J. Mar. Sci. Technol.}, 20(3), 559--578.

\bibitem[{Lamraoui and Qidan(2019)}]{lamraoui2019path}
Lamraoui, H.C. and Qidan, Z. (2019).
\newblock Path following control of fully-actuated autonomous underwater vehicle in presence of fast-varying disturbances.
\newblock \emph{Appl. Ocean Res.}, 86, 40--46.

\bibitem[{Lekkas and Fossen(2013)}]{lekkas2013line}
Lekkas, A.M. and Fossen, T.I. (2013).
\newblock Line-of-sight guidance for path following of marine vehicles.
\newblock In \emph{Advanced in Marine Robotics}, volume~5, 63--92. LAP Lambert Academic Publishing.

\bibitem[{Liang et~al.(2017)Liang, Qu, Hou, and Zhang}]{liang2017three}
Liang, X., Qu, X., Hou, Y., and Zhang, J. (2017).
\newblock Three-dimensional path following control of underactuated autonomous underwater vehicle based on damping backstepping.
\newblock \emph{Int. J. Adv. Robot. Syst.}, 14(4), 1729881417724179.

\bibitem[{Naeem et~al.(2003)Naeem, Sutton, Ahmad, and Burns}]{naeem2003review}
Naeem, W., Sutton, R., Ahmad, S., and Burns, R. (2003).
\newblock A review of guidance laws applicable to unmanned underwater vehicles.
\newblock \emph{J. Navigation}, 56(1), 15--29.

\bibitem[{Ribas et~al.(2015)Ribas, Ridao, Turetta, Melchiorri, Palli, Fern{\'a}ndez, and Sanz}]{ribas2015auv}
Ribas, D., Ridao, P., Turetta, A., Melchiorri, C., Palli, G., Fern{\'a}ndez, J.J., and Sanz, P.J. (2015).
\newblock {I-AUV} mechatronics integration for the {TRIDENT FP7} project.
\newblock \emph{IEEE/ASME Trans. Mechatron.}, 20(5), 2583--2592.

\bibitem[{Sahoo et~al.(2019)Sahoo, Dwivedy, and Robi}]{sahoo2019advancements}
Sahoo, A., Dwivedy, S.K., and Robi, P. (2019).
\newblock Advancements in the field of autonomous underwater vehicle.
\newblock \emph{Ocean {E}ng.}, 181, 145--160.

\bibitem[{Yan and Wang(2012)}]{yan2012model}
Yan, Z. and Wang, J. (2012).
\newblock Model predictive control for tracking of underactuated vessels based on recurrent neural networks.
\newblock \emph{IEEE J. Oceanic Eng.}, 37(4), 717--726.

\bibitem[{Yao et~al.(2020)Yao, Wang, Wang, and Zhang}]{yao2020path}
Yao, X., Wang, X., Wang, F., and Zhang, L. (2020).
\newblock Path following based on waypoints and real-time obstacle avoidance control of an autonomous underwater vehicle.
\newblock \emph{Sensors}, 20(3), 795.

\bibitem[{Yu et~al.(2017)Yu, Xiang, Lapierre, and Zhang}]{yu2017nonlinear}
Yu, C., Xiang, X., Lapierre, L., and Zhang, Q. (2017).
\newblock Nonlinear guidance and fuzzy control for three-dimensional path following of an underactuated autonomous underwater vehicle.
\newblock \emph{Ocean {E}ng.}, 146, 457--467.

\bibitem[{Zeng et~al.(2015)Zeng, Lian, Sammut, He, Tang, and Lammas}]{zeng2015survey}
Zeng, Z., Lian, L., Sammut, K., He, F., Tang, Y., and Lammas, A. (2015).
\newblock A survey on path planning for persistent autonomy of autonomous underwater vehicles.
\newblock \emph{Ocean {E}ng.}, 110, 303--313.

\bibitem[{Zhang et~al.(2019)Zhang, Liu, Luo, and Yang}]{zhang2019mpc}
Zhang, Y., Liu, X., Luo, M., and Yang, C. (2019).
\newblock {MPC}-based 3-{D} trajectory tracking for an autonomous underwater vehicle with constraints in complex ocean environments.
\newblock \emph{Ocean {E}ng.}, 189, 106309.

\bibitem[{Zhang et~al.(2024)Zhang, Li, Wang, Wang, and Mu}]{zhang2024three}
Zhang, Y., Li, S., Wang, W., Wang, S., and Mu, R. (2024).
\newblock Three-dimensional path following control of underactuated autonomous underwater vehicles with nonzero roll dynamics: A novel line-of-sight-guided approach.
\newblock \emph{Int. J. Robust Nonlinear Control,}, 34(8), 4959--4977.

\bibitem[{Zhou et~al.(2013)Zhou, Tang, Zhang, Jiao et~al.}]{zhou2013spatial}
Zhou, J., Tang, Z., Zhang, H., Jiao, J., et~al. (2013).
\newblock Spatial path following for {AUVs} using adaptive neural network controllers.
\newblock \emph{Math. Probl. Eng.}, 2013.

\end{thebibliography}

\end{document}